\documentclass[aps,prl,reprint,superscriptaddress, longbibliography]{revtex4-2}
\usepackage{graphicx}
\usepackage{natbib}
\usepackage{xcolor}
\usepackage{verbatim}

\definecolor{pink}{RGB}{255,105,180}

\usepackage{graphicx}   

\usepackage{url}        

\usepackage{amsmath}    

\usepackage{siunitx}
\DeclareSIUnit\angstrom{\text{\AA}}
\usepackage[normalem]{ulem}

\def \ILfrac#1#2{{#1/#2}} 

\begin{document}

	\title{Observation of a magnetic shift in neutron whispering-gallery states}

    \newcommand{\authorspacing}{\vspace{0.7em}}
    
    \author{K. Schreiner} 
    \affiliation{Marietta Blau Institute, Austrian Academy of Sciences, Vienna, Austria}
    \affiliation{Laboratoire Kastler Brossel, Sorbonne Universite, CNRS, ENS-Universite PSL, College de France, Paris, France}
    \affiliation{Vienna Doctoral School in Physics, Vienna, Austria}
    
    \author{J. A. Pioquinto} 
    \affiliation{University of Virginia, Charlottesville, Virginia, USA}
    \affiliation{Institut Max von Laue -- Paul Langevin, Grenoble, France}
    
    \author{S. Bae\ss{}ler} 
    \affiliation{University of Virginia, Charlottesville, Virginia, USA}
    \affiliation{Oak Ridge National Laboratory, Oak Ridge, Tennessee, USA}
    
    \author{V. V. Nesvizhevsky} 
 \affiliation{Institut Max von Laue -- Paul Langevin, Grenoble, France}
    
    \author{P. Cladé}
    \affiliation{Laboratoire Kastler Brossel, Sorbonne Universite, CNRS, ENS-Universite PSL, College de France, Paris, France}

     \author{P. Crivelli} 
     \affiliation{Institute for Particle Physics and Astrophysics, ETH, Zurich, Switzerland}
    
    \author{R. Cubitt}
 \affiliation{Institut Max von Laue -- Paul Langevin, Grenoble, France}
    
    \author{J. Guyomard}
    \affiliation{Laboratoire Kastler Brossel, Sorbonne Universite, CNRS, ENS-Universite PSL, College de France, Paris, France}

   \author{C. Killian}
   \affiliation{Marietta Blau Institute, Austrian Academy of Sciences, Vienna, Austria}

     \author{U. Miniotaite} 
 \affiliation{Institut Max von Laue -- Paul Langevin, Grenoble, France}
    
    \author{F. Nez}
    \affiliation{Laboratoire Kastler Brossel, Sorbonne Universite, CNRS, ENS-Universite PSL, College de France, Paris, France}

    \author{E. Perry} 
 \affiliation{Institut Max von Laue -- Paul Langevin, Grenoble, France}
    
    \author{S. Reynaud}
    \affiliation{Laboratoire Kastler Brossel, Sorbonne Universite, CNRS, ENS-Universite PSL, College de France, Paris, France}
    
    \author{T. Saerbeck}
 \affiliation{Institut Max von Laue -- Paul Langevin, Grenoble, France}

     \author{L. Shen} 
     \affiliation{University of Virginia, Charlottesville, Virginia, USA}
 \affiliation{Institut Max von Laue -- Paul Langevin, Grenoble, France}

     \author{K. Stinson} 
     \affiliation{University of Virginia, Charlottesville, Virginia, USA}
 \affiliation{Institut Max von Laue -- Paul Langevin, Grenoble, France}

         \author{J. Vivier} 
 \affiliation{Institut Max von Laue -- Paul Langevin, Grenoble, France}

\author{A. Vorobiev}
\affiliation{Department for Physics and Astronomy, Uppsala University, Uppsala, Sweden}
 \affiliation{Institut Max von Laue -- Paul Langevin, Grenoble, France}

     \author{J. Walsh} 
 \affiliation{Institut Max von Laue -- Paul Langevin, Grenoble, France}
      \affiliation{KTH Royal Institute of Technology, Department of Applied Physics, Stockholm, Sweden}
     
    \author{E. Widmann}
    \affiliation{Marietta Blau Institute, Austrian Academy of Sciences, Vienna, Austria}
    
    \author{P. Yzombard}
    \affiliation{Laboratoire Kastler Brossel, Sorbonne Universite, CNRS, ENS-Universite PSL, College de France, Paris, France}

     \author{A. Zhao} 
 \affiliation{Institut Max von Laue -- Paul Langevin, Grenoble, France}
     \affiliation{University of Virginia, Charlottesville, Virginia, USA}

\begin{abstract}

We developed experimental and theoretical methods to create and describe high-resolution whispering-gallery interference patterns and show the feasibility of measuring their small shifts by external interactions. Such experiments can be used to search for extra fundamental interactions, time parity violations, nonzero electric charges, precisely measuring the gravitational properties of (anti)matter, parity violations, neutron polarizability, quantum reflection, surface state effects, etc. Here, we measure a magnetic shift of such a neutron pattern 
$\delta g/g\sim (6.5 \pm 0.8_{st} \pm 1.2_{sys}) \cdot 10^{-4}$, and our sensitivity to spin-dependent differences between the scattering lengths was $\delta b_n/b_n \sim 10^{-4}$.

\end{abstract}

\maketitle

\emph{Introduction}. The use of interference techniques for ultra-sensitive or precise studies is a well-known method \cite{Bunch2004,Hariharan2007}. The quantum bouncer \cite{Breit1928} (invented at the dawn of quantum mechanics) provides fairly complex interference patterns. Two realizations were discovered with massive particles: Gravitational Quantum States \cite{Nesvizhevsky2002} (GQS) (a particle bouncing over a flat mirror in a gravitational field) and Whispering Gallery States \cite{Nesvizhevsky2010} (WGS) (a particle bouncing over the surface of a curved mirror), generalized these phenomena to other light neutral particles \cite{Voronin:2011,Voronin2012, Nesvizhevsky2026} and proposed a new method \cite{Nesvizhevsky2026} of measuring small shifts in WGS and GQS interference patterns to conduct ultra-sensitive or precise measurements of various small interactions and phenomena. 

The particles include neutrons ($n$), (anti)atoms, muonium ($Mu$) and positronium ($Ps$) \cite{Nesvizhevsky2026}. Optical $n$-nuclei potential \cite{Fermi1936} of the surface reflects $n$. Van der Waals -- Casimir-Polder potential \cite{Friedrich2002} reflects charged particles via Quantum Reflection (QR). GQS and WGS are quantum sensors for extra fundamental interactions \cite{MoodyWilczek1984,Dobrescu2006,Antoniadis2011, Fadeev2019}, spatial and time reversal invariance violation \cite{Flambaum2022, vesna2021podd}, non-zero 
electric charges \cite{Baumann1988}. Gravitational properties of matter \cite{Takahashi2026} or antimatter \cite{Anderson2023}, $n$ polarizability, QR of matter \cite{Yu1993} or antimatter \cite{Voronin2005}, surface potential and state effects could be measured precisely using this method. 

GQS were first observed \cite{Nesvizhevsky2003} using Ultra Cold Neutrons (UCN) \cite{Lushchikov1969} following the idea proposed in \cite{Lushchikov1978} and the method developed in \cite{Nesvizhevsky2000nima}. 
The results are described in \cite{Nes2005, Voronin2006, Westphal2007}. Several collaborations developed this method (qBounce \cite{Jenke2011,Sedmik2019}, Tokyo \cite{Ichikawa2014}, GRANIT \cite{Baessler2011}) and used it for some BSM searches (e.g., \cite{Jenke2014,Cronenberg2018}). However, available UCN velocities and mirror sizes limit the precision and sensitivity. To overcome these problems, we pursue GQS studies with atoms \cite{Killian2023,Killian2024}, develop UCN sources \cite{Sakhiyev2025} and ultracold hydrogen sources \cite{Ahokas2022,Semakin2025} with a higher density, and develop methods of long storage of particles in GQS \cite{Nesvizhevsky2020mgt}.

In contrast to GQS, WGS can provide both larger statistics and many quasi-classical bounces. The method was proposed in \cite{Nesvizhevsky2008} and tested in \cite{Cubitt2009}. The phenomenon was observed in \cite{Nesvizhevsky2010} and theoretically described in \cite{Nesvizhevsky2010njf}. Its potential to search for extra fundamental interactions was noted in \cite{Nesvizhevsky2011crp}. Advantages of using pulsed $n$ sources are noted in \cite{Ichikawa2025}. Overviews of GQS and WGS research are available \cite{Baessler2009, Nesvizhevsky2010ufn}. Whispering galleries were observed for a variety of other waves in 115 years
\cite{Rayleigh1910, Rayleigh1914, Raman1921, Budden1962, Mabuchi1994, Liu1997, Oraevsky2002, Hyun2008, Chiasera2010, Mendis2010, Reecht2013}.
In the last decade, we developed methods to create high-resolution $n$ interference patterns and describe them with high precision and reliability (some complementary techniques are mentioned in the Acknowlegment). 

Here, we combine these developments and show the feasibility of measuring small shifts in the interference patterns using the magnetic shift of a $n$ WGS. The new method complements well-tested methods for measuring the wave functions of quantum states \cite{Nesvizhevsky2002,Ichikawa2014} and the energies of resonant transitions between quantum states \cite{Nesvizhevsky2006,Kreuz2009,Jenke2011,Pignol2014}. It eliminates some systematic effects and increases statistics. Our data analysis procedure is applicable to future analogous measurements with $n$, (anti)atoms, $Mu$, and $Ps$.

\begin{figure*}[ht]
    \centering
    \includegraphics[width=0.7\linewidth]{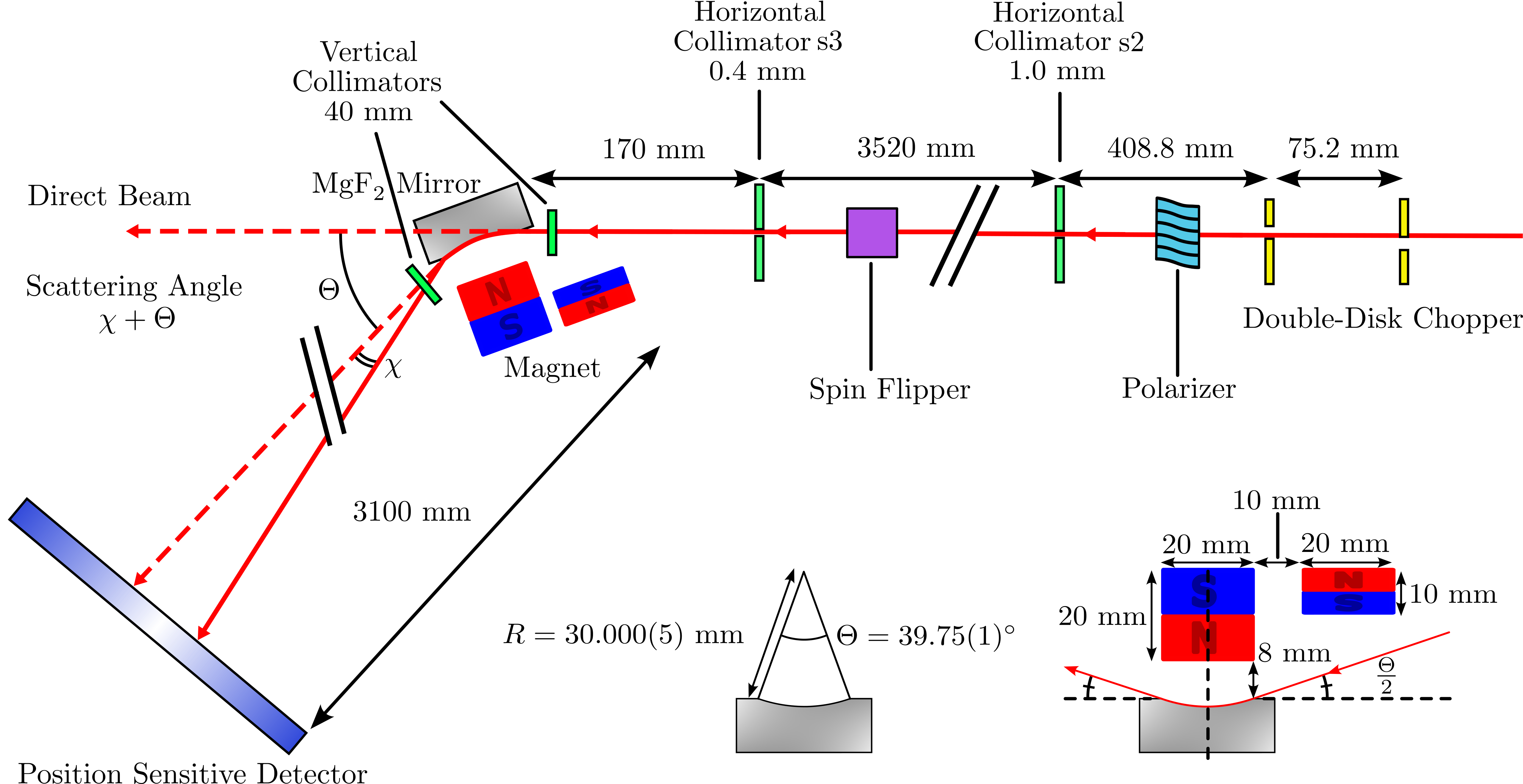}
    \caption{The experimental setup, top view, described in text.}
    \label{fig:setup}
\end{figure*}

\emph{Experimental Setup}. A schematic of the setup of the experiment on the D17 beamline \cite{Saerbeck2022} at the Institut Laue Langevin (ILL) in Grenoble is shown in Fig.~\ref{fig:setup}. The $n$ beam is chopped, then polarized, and depending on the desired polarization state, spin flipped. It impinges on the curved mirror, after being shaped by the slits $s_2$ and $s_3$. Its width is $\sim$\qty{0.4}{\milli\meter}
  ~(much larger than WGS modes sizes of 
$15-150$~nm \cite{Nesvizhevsky2010njf}), height (out-of the plane) is 4~cm, and horizontal angular divergence is $\sim$0.4~mrad (significantly less than angular acceptance of WGS modes of $\sim$2.5-12.5~mrad)~ \cite{Nesvizhevsky2010njf}. Thus, we treat $n$ as plane waves.

We used a $\mathrm{MgF}_2$ single-crystal mirror with cylindrical cutout, an opening angle of $\Theta = 39.75(1)^\circ$ and a curvature radius of $R = 30.000(5)$~mm, 
fixed in a Teflon holder such that the symmetry axis of the cylinder was upright.

Several stacks of magnets provide a strong average field gradient of \qty{20}{\tesla\per\meter} at the curved mirror surface and \qty{4}{\tesla\per\meter} at the incoming beam. They induce a spin-dependent shift in the interference pattern.

The outgoing beam impinges on a $^3\mathrm{He}$ position-sensitive detector (PSD) far behind the mirror, which records the interference pattern formed by $n$ as a function of time and coordinate. 
We summed over the vertical coordinate and obtained a one dimensional (1-D) spatial interference pattern with the spatial resolution of \qty{2.2}{\milli\meter} in the horizontal plane. The detection time $t_{ToF}$ is converted into $n$ wavelength $\lambda_n$ with the de Broglie relation $\lambda_n = \frac{h}{m}\frac{t_{ToF}-t_0}{D_{ToF}}$. 
The time-of-flight (ToF) distance $D_{ToF}$ and timing offset $t_0$ are initially determined by the setup's and chopper-triggering system geometry, and more precisely by fitting the measured patterns with a theoretical model, which yields $D_{ToF}=\SI{7252(3)}{\milli\meter}$ and $t_0 = \SI{25(1)}{\micro\second}$.

\emph{Theory}. $n$ propagating near the mirror can be confined to the vicinity of the surface. In a reference frame co-rotating with a $n$, the $n$ will experience a centrifugal acceleration $g_\text{WGS}=v_{\parallel}^2/R=h^2/m^2R\lambda_n^2$ in the radial direction, where $v_{\parallel}$ is the $n$ velocity parallel to the surface. The $n$ position with respect to the surface is indicated as $z = R - r$ where $r$ is the radial coordinate with its origin on the cylinder axis.

For $n$ with a low velocity perpendicular to the surface, $v_{\perp}$, their interaction with the gallery can be described as a potential step of height $U_0$ ($n$-nuclei optical potential \cite{Fermi1936} of the mirror material). The surface roughness effect is modeled with a smoothed potential step of the form $U_0/(1+\exp(z/a))$ \cite{PhysRev.95.577} with $a = 5.3 \text{ \AA}$. 

The magnet fixed above the mirror introduces a spin-dependent potential  $-\vec{\mu_n}\cdot\vec{B}$ from a magnetic field $\vec{B}$ with a gradient of $\partial_z |\vec{B}(z)|$, where the $n$ magnetic moment $\vec{\mu}_n$ is aligned with the $\vec{B}$ direction provided the rate of variation of the field is much slower than its Larmor frequency \cite{Majorana1932, vladimirskii1961magnetic}. Then, 
$U_B^\pm=-\vec{\mu_n}\cdot\vec{B}\approx \mp \mu_n |\vec{B}|$, where the spin-up 
(defined as 'spin aligned with magnetic field') 
 is denoted as $U_B^+$ and the spin-down as $U_B^-$.
The potential experienced by $n$ in a WGS is then \cite{Nesvizhevsky2010njf}:
\begin{equation}
    U^\pm(z) =mg_{\mathrm{WGS}}z + \frac{U_0}{1 + \exp(z/a)} \mp \mu_n |\vec{B}(z)|.
    \label{eq: Potential}
\end{equation}
The evolution of its wave function under Eq. \ref{eq: Potential} is determined by the time dependent Schrödinger equation, the solution of which can be approximated as a discrete sum of quasi-stationary states $\psi_n$ with complex energy levels $E_n$ such that 
    $\psi(z,t)\approx \sum_n c_n \psi_n(z) e^{-\frac{iE_n t}{\hbar}}$.
We find it to be accurate enough compared to numerical simulations \cite{Pioquinto2025WhisperingGallery, pioquinto2026WGtheory}. 

\begin{figure}
    \centering
    \includegraphics[width=1\linewidth]{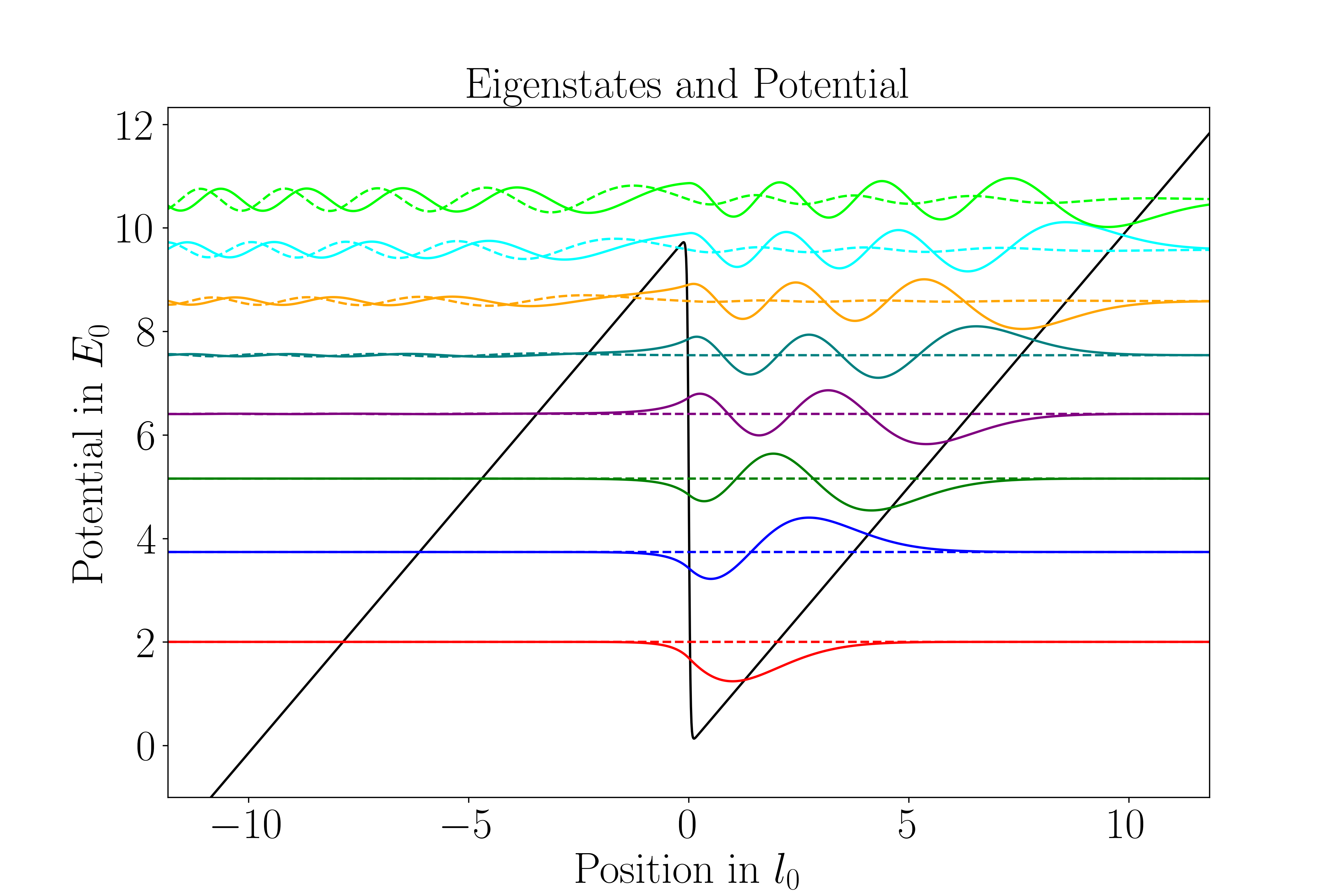}
    \caption{The black line shows the centrifugal plus the $n$-nuclei optical potential, with the mirror surface at $z=0$. The colored curved lines are the real (solid) and imaginary (dashed) parts of the eigen states $\psi_n$ as a function of position above the surface, with an offset on the vertical axis corresponding to their energy $E_n$.
    Each quantity is plotted in the characteristic length $l_0 
= \sqrt[3]{\frac{ \hbar^2R}{2M^2v^2}}$ and energy $E_0 = mg_\text{WGS}l_0$ units, respectively. Here, $ l_0 = \qty{39.3}{\nano\meter}$ and $E_0 = \SI{13.4}{\nano\electronvolt}$ evaluated for $\lambda_n = \qty{4}{\angstrom}$. }
    \label{fig: Eigenstate}
\end{figure}

To account for the surface roughness, logarithmic perturbation theory \cite{leung1998qnm} is used to find first order corrections to $\psi_n, E_n$, where Eq. \ref{eq: Potential} with $a = 0$ produces the unperturbed
solutions. Examples of the eigenstates and the potential can be seen in Fig. \ref{fig: Eigenstate}. 

The PSD is placed far from the mirror, thus the measured interference pattern is proportional to the magnitude squared of the wave function in momentum space at the end of the gallery \cite{Nesvizhevsky2025MagneticShift}   $\tilde{\psi}(p) = 1/\sqrt{2\pi\hbar}\int_{-\infty}^\infty \psi(z, T)\exp{(-ipz/\hbar)}dz$
with propagation time $T = R\Theta/v$. 

\emph{Effective Wavelength Shifts}. The magnetic field effect can be effectively described as a spin-dependent shift of the interference pattern in wavelength space $\delta \lambda_n$. A small spin-dependent change $\delta g^\pm_\text{WGS}$ leads to
\begin{equation}
    \delta \lambda_n^\pm = -\frac{1}{2}\frac{m^2R}{h^2} \delta g^\pm_\text{WGS}\lambda_n^3.
    \label{eq: Wavelength Shift}
\end{equation}
with $+$ and $-$ indicating spin and down, respectively.

We expand $U_B^\pm\approx \mp\mu_n (|\vec{B}||_{z=0}  + \partial_z |\vec{B}||_{z=0} \times z )$ along the gallery surface.

The term linear in $z$ acts as a spin-dependent force which adds to the shift in centrifugal acceleration $\mp \mu_n\partial_z |\vec{B}||_{z=0}/m$.
A $\lambda_n$ shift also arises from the  constant term $\mp\mu_n |\vec{B}||_{z=0}$ in the expansion of $U_B^\pm$. This spin-dependent constant adds a shift to the $n$'s kinetic energy of $\delta E_\pm = \pm \mu_n |\vec{B}|\big|_{z=0}$,
which is
related to $ \delta \lambda_n^{\pm}$ by
$E = \frac{h^2}{2m\lambda_n^2} \rightarrow \delta \lambda^\pm_n = -\delta E_\pm\frac{m}{h^2}\lambda_n^3$.
This has the same form as Eq. (\ref{eq: Wavelength Shift}), so it behaves as another shift in $g_\text{WGS}$. 
Taken together:
\begin{equation}
    \delta g_\text{WGS}^{\pm} =  \mp \frac{\mu_n}{m}\partial_z |\vec{B}|\big|_{z=0} + 2\frac{\delta E_\pm}{m R}.
    \label{eq: shift g total center}
\end{equation}

\begin{figure}
    \centering
    \includegraphics[width=\columnwidth]{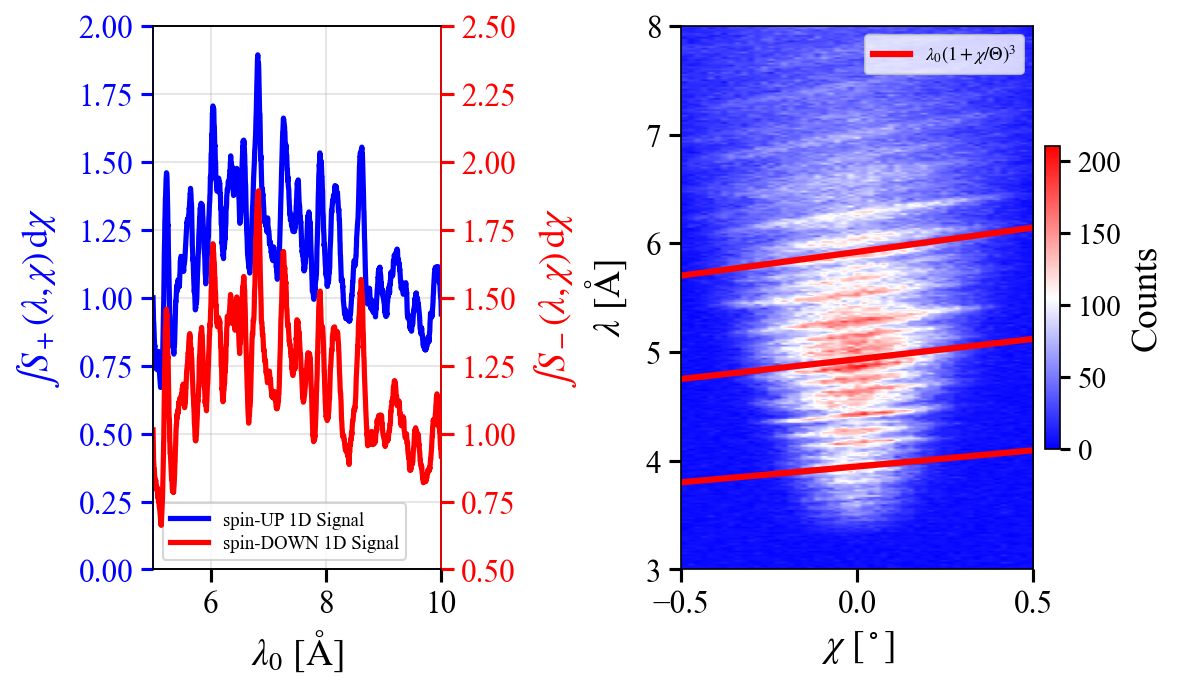}
    \caption{The 1-D signals for both polarizations (left) obtained from the 2-D signal (right). The shift is too small to see by eye. We represent the $n$ data in $\chi$ and $\lambda_n$ as $S(\lambda_n, \chi)$. The red lines (right) show 
    several directional integration lines. 
    }
    \label{fig:lines}
\end{figure}

Two other effects contribute to $\delta g_\text{WGS}^{\pm}$, the deflection before the beam 
enters the mirror system, and after it exits. 
In the semi-classical approximation, the wave function can be treated as if it is in free fall in an acceleration field of strength $g_\text{WGS}$ 
\cite{Breit1928,Dufour2015}. 
In momentum space 
it is $\tilde{\psi}(p/p_0 + T/t_0)$,
where $p$ is the momentum in the direction perpendicular to the surface, $p_0 = \hbar/l_0$ and $t_0 = \hbar/E_0$ 
are the characteristic momentum scale and
time scales. The inclined interference lines (see Fig. \ref{fig:lines}) can be thought of as the lines of constant phase
$\phi = p/p_0 + T/t_0$.
Requiring $\text{d}\phi = \frac{\partial \phi}{\partial p} \big|_v\text{d}p + \frac{\partial \phi}{\partial v}\big|_p \text{d}v = 0$ and using scattering angle $\chi = p/mv$, we find  
\begin{equation}
    \lambda_n = \lambda_{n_0} \left(1+\frac{\chi}{\Theta}\right)^3 ,
    \label{eq: line inclination}
\end{equation}
where $\lambda_{n_0}$ is $\lambda_n$ 
for $\chi=0$. This line shape agrees well with the data, see Fig. \ref{fig:lines}. 

With Eq. (\ref{eq: line inclination}) established, we see that a small change of 
$\delta\chi$ results in a shift in $\lambda_n$ 
\begin{equation}
    \delta \lambda_n = \frac{3}{1+\ILfrac{\chi}{\Theta}} \frac{\delta \chi}{\Theta}\lambda_n \approx3\frac{\delta\chi}{\Theta}\lambda_n  ,
    \label{Eq: shift angle}
\end{equation}
for $\chi \ll \Theta$. 
Eq. \ref{Eq: shift angle} 
equivalently describes the effect of a change in $\alpha$, 
which is justified by the classical relationship $p = mv\chi = mv\alpha - m g_\text{WGS}t$, where $v\alpha$ is the initial velocity in the direction perpendicular to the mirror.
Thus, a change $\delta \alpha$ results in a change $\delta \chi$.

The expected shifts $\delta \chi$ or $\delta \alpha$ can be found by calculating the change in the $n$ velocity 
as they pass through
the acceleration field $\vec{a}(\vec{r}(t))$ along their trajectory $\vec{r}(t)$ when entering or leaving the gallery. The calculation of the field gradient (thus, the acceleration field) is 
discussed in the Appendix (see Fig. \ref{B_fields} and the text around it). 

The incidence angle 
is $\alpha = v_\perp/v_\parallel$, where $v_\perp$ and $v_\parallel$ are the velocities perpendicular and parallel to the mirror surface at the entrance.
Small changes in $v_\perp$ and $v_\parallel$
induce a change in 
$\delta \alpha \approx \delta v_\perp/v$
neglecting the second order contribution from changes in $v_\parallel$ and considering that $v_\parallel \approx v$ since $v_\perp \ll v_\parallel$. The change in $v_\perp$ 
is $\delta v_\perp = \int_{t_0}^{t_1} a_\perp(t) dt \approx \int_{l_0}^{l_1} a_\perp(l) dl/v = \delta W_\perp/mv$ 
where we change coordinates from ToF $t$ to path length $l = v_\parallel t\approx vt$, which assumes the changes in the trajectory and 
$v_{\parallel}$ 
are small, which we find to be true for our field gradient. 
Then $ \delta \alpha  \approx \ILfrac{\delta W_\perp}{mv^2}= (\ILfrac{m^2}{h^2}) \lambda_n^2 \delta W_\perp/m$,
and
Eq. (\ref{Eq: shift angle}) results in a  
shift $\delta \lambda_n^\pm =\pm (\ILfrac{m^2}{h^2})(\ILfrac{3\delta W_\perp}{m\Theta}) \lambda_n^3$,
which formally matches Eq. (\ref{eq: Wavelength Shift}). The sign is determined by the spin of $n$, which determines the direction of deflection. By combining each contribution to $\delta g_\text{WGS}^{\pm}$, the overall magnetic shift is 
\begin{equation}
    \delta g_\text{WGS}^{\pm} =  \mp\frac{\mu_n}{m}\partial_z |\vec{B}|\big|_{z=0} + 2\frac{\delta E_\pm}{m R} \mp \frac{6}{mR\Theta} (\delta W^{\text{in}}_\perp + \delta W^{\text{out}}_\perp) ,
    \label{eq:OverallShift}
\end{equation}
where $\delta W_\perp^{\text{in},\text{out}}$ refers to the deflection of the incoming or outgoing beam.

Since we measured interference patterns of two spin states, we will extract the shift between these two patterns $\delta g_{\mathrm{WGS}} = \delta g^+_{\mathrm{WGS}} - \delta g^-_{\mathrm{WGS}}$, or equivalently $\delta \lambda_n = \delta\lambda_n^+ - \delta\lambda_n^-$.

\emph{Measurements}. 
This study was conducted in two beam-times with increasing sensitivity and reliability. In both cases, 
we observed clear interference fringes for WGS formation on $\mathrm{MgF}_2$. In 2025, we opened the slit $s_3$ to 
\qty{0.4}{\milli\meter} to reduce misalignment effects due to thermal drifts \cite{Nesvizhevsky2025MagneticShift}, 
reduced run durations (1 min) and adopted a scheme $+--+-++-$ to account for linear and quadratic drifts.
The count rate became statistically 
stable 
and only changed by an offset due to background caused by the occasional change in the shutter-position of neighboring experiments. 

The resulting 1-D and 2-D interference patterns are shown in Fig. \ref{fig:lines}, the expected theoretical and measured (using both the 1-D and 2-D methods) values of the shift are summarized in Table \ref{tab:results} and commented in the caption.
Details and methods are presented in the Appendix.

\begin{table}[h]
\centering
\begin{tabular}{lcc}
\textbf{Method} &
\textbf{$\delta g_{\mathrm{WGS}}$ (\SI{}{\meter\per\second\squared})} & $\delta g_\text{WGS}/g_\text{WGS}^{\lambda_n=\qty{10.3}{\angstrom}}$\\ \hline
\hline
Theory & $3.2(0.2)\times 10^3$ & $6.7(0.4) \times 10^{-4}$ \\ 
1-Dimensional fit  & $3.9(1.0)\times 10^3$ & $8.0(2.0) \times 10^{-4}$ \\ 
2-Dimensional fit  & $3.7(0.4)\times 10^3$  & $7.5(0.8) \times 10^{-4}$ \\ 
\end{tabular}
\caption{
Three methods of evaluation of the acceleration shift $\delta g_{\mathrm{WGS}}$ induced by the magnetic field: theory, 1-D and 2-D analysis. 
Absolute shifts, in $\mathrm{m/s^2}$, and relative shifts normalized to $\lambda_n=\qty{10.3}{\angstrom}$ providing the best sensitivity 
(see Fig. \ref{fig:sensitivity} in the Appendix and the related text). All three estimations provide consistent results; the 2-D analysis provides a better precision that the simplified 1-D analysis. Both 1-D and 2-D analysis show non-zero results.}
\label{tab:results}
\end{table}

As a by-path result of this short run, our sensitivity to parity-violating differences between scattering lengths was 
$\delta b_n/b_n =\delta U_0/U_0 \sim 10^{-4}$ by including a spin-dependent offset to the $n$-nuclei 
optical potential of the mirror as a parameter in our fit of the data (using Eq. \ref{eq: Potential}).

\emph{Conclusion}. 
We have developed a method to precisely measure small accelerations in WGS,
as well as
a suitable analysis strategy. The method was demonstrated with
neutrons at the D17/ILL instrument using a
cylindrical $\mathrm{MgF}_2$ mirror. 
We succeeded in measuring 
a magnetic shift in WGS interference patterns.  A relative magnetic acceleration contribution was $(7.5 \pm 0.8_{stat} \pm 0.4_{syst}) \cdot 10^{-4}$ (normalized at $\lambda_n = \qty{10.3}{\angstrom}$).

The motivation for developing this method spans from testing fundamental invariance violations, short range interactions, and 
electric charge to testing fundamental effects like quantum reflection and testing gravitational properties of antimatter. 
Possible experiments are described in detail in \cite{Nesvizhevsky2026}; they  
can also be performed with other neutral 
particle species.
In the near future, we are going to perform a set of measurements using cold hydrogen and deuterium atoms at MBI in Vienna.

\emph{Acknowledgments}. We thank all the 
GRANIT and GRASIAN collaborators
as well as the D17 and NPP/ILL staff.
JP and SB are supported by the NSF under grant PHY-2412782. JP was supported 
by The Beitchman Award for Innovative Graduate Student Research in Physics in honor of Robert V. Coleman and Bascom S. Deaver, Jr. KS, EW, JP and VV are supported by AMADEUS under grant ‘Measurements of whispering gallery states (WGS) of neutrons and atoms’.
Mirror characterization measurements were conducted by the MML/ESRF: 3D surface profile measurements made using a Veeco NT9300 Micro-interferometer and a Keyence confocal microscope VK-X1100 by F. Perrin.
Our magnets were mapped using a magnetometer on a robotic arm by D. Jullien at the NOP/ILL. 

\emph{Author Contribution Statement}.
Data curation: KS, JP, SB, VN, RC, FN, TS, KS, AV, EW;
Conceptualization: KS, JP, SB, VN;
Software: KS, JP;
Formal analysis: KS, JP, PC, UM, EP, SR, LS, JV, JW;
Supervision: SB, VN, FN, EW; 
Funding: JP, SB, VN, EW;
Validation: KS, JP, SB, VN;
Investigation: KS, JP, SB, VN, FN, TS, EW; Visualization: KS, JP;
Methodology: KS, JP, SB, VN, PC, RC, FN, SR, TS, AV;     
Writing the original draft: KS, JP, SB, VN;
Writing review and editing: KS, JP, SB, VN, JG, UM, EP, JW, EW.

\emph{Data availability statement}. $n$ experiments were performed using D17 \cite{Saerbeck2022}, PF1B \cite{Abele2006} and SuperADAM \cite{Devishvili2018} instruments at the ILL: VN et al (2016) doi:10.5291/ILL-DATA.3-15-85; (2018) doi:10.5291/ILL-DATA.3-15-90; (2018) doi:10.5291/ILL-DATA.3-15-92; (2019) doi:10.5291/ILL-DATA.3-15-92; (2024) doi:10.5291/ILL-DATA.3-15-101; (2025) 10.5291/ILL-DATA.DIR-388.

\bibliography{Refs}

\appendix

\newpage

\emph{Data Analysis}. Before applying the analysis procedure to real data, we verified it with simulations. The aim was to extract the known value of the extra acceleration $\delta g_{\mathrm{WGS}}$ from the induced shift $\delta\lambda_n$ in the respective interference patterns $S(\lambda_n,\chi)$. Two procedures were developed for this to ensure reliable results. They were tested by performing a blind analysis of several simulations convoluted with realistic experimental resolution parameters: the time resolution given by the chopper and spatial resolution of the pixels on the PSD.

\emph{1-D Signal Analysis}. Here, 2-dimensional (2-D) interference patterns $S(\lambda_n, \chi)$ (Fig. \ref{fig:lines} right) are reduced to one-dimensional (1-D) curves (Fig. \ref{fig:lines} left), and the shift parameter $\delta\lambda_n$
is extracted from the fitting of the 1-D curves. The magnetic gradient field causes a spin-dependent shift in the 2-D signal $S_\pm(\lambda_n, \chi)$. 
We denote $S_{\pm}$ for the signal with polarized neutrons.
We assume that we can find $\delta \lambda_n$ such that $S_-({\lambda_n} - \delta{\lambda_n}^{1D}/2, \chi) = S_{\delta_g}(\lambda_n + \delta{\lambda_{n}}^{1D}/2, \chi)$ and $S_-(\lambda_n, \chi) = S_{\delta_g}(\lambda_n, \chi)$. The 1-D  
curve is obtained by integrating 
the 2-D 
signal $S_\pm(\lambda_n,\chi)$ along the interference lines, described by Eq. (\ref{eq: line inclination}). 
In the small angle approximation, these interference lines 
are inclined lines in $S(\lambda_n, \chi)$ , which have maximum intensity around $\chi = 0$.
The substitution $S_\pm (\lambda_{n_0} ) =  \int S_\pm(\lambda_{n_0} \cdot \left( 1 + \chi/\theta\right)^3, \chi) \mathrm{d \chi}$ transforms our shift condition into $S_{UP}(\lambda_{n_0}- \delta{\lambda_{n_0}}^{1D}/2)=S_{DOWN}(\lambda_{n_0} + \delta\lambda_{n_0}^{1D}/2)$. 
Since the B-fields present in our setup 
cause a $\chi$ shift of the direct beams of the two different polarizations on the PSD in a Stern-Gerlach effect.
To separate this effect from our magnetic shift, we
modify the $\chi$ in the analysis, to make sure only the shift $\delta\lambda_n$ will determine the result of the fitting procedure. We therefore determine the center-of-mass of $S_\pm({\lambda_n}, \chi)$, and use $\chi^{\pm}$ such that $\chi^\pm = 0$ at the center of mass of the respective polarization orientations.

In a Monte-Carlo method described below, we then evaluate the 
$\delta\lambda_n$ which minimizes 

\begin{equation}
    \Sigma_{\lambda_n}|S_-(\chi) - S_{\delta \lambda_{n_0}^{1D}}(\chi)|^2
    \label{resid}
\end{equation}
assuming again that $S_{\delta \lambda_{n_0}^{1D}} = S_+$.
The $\delta \lambda_{n_0}$ is found via a fitting procedure performed on \ref{resid}, where the fitting space is $\lambda_{n0}$. The uncertainty is then determined via $N-$fold repetition of the fit on the data with additional random Poissonian error added to the sum of all datasets collected throughout the beamtime.
This gives statistical meaning to the uncertainty of the value of $\delta \lambda_{n0}^{1D}$ obtained with this analysis method. Via Eq. \ref{Eq: shift angle} we relate this to the shift of $\delta_g$ shown in Table \ref{tab:results}. 

\emph{2-D Signal Analysis}. We also directly fit the spin-down and spin-up interference patterns in 2-D with the theoretical model detailed in the Theory section. 
The fit function corresponds to the probability current at the exit of the mirror $j(x, \lambda_n, \alpha)$, where $x$ is the position measured on the PSD, 
and $\alpha$ is the incidence angle of the beam at the entrance of the mirror. To account for the shape of the beam spectra, $n(\lambda_n)$, we multiply $j$
by the spectra measured during the experiment when the mirror was placed completely out of the beam. This product is then convolved with the resolution function of the instrument $\mathcal{R}(x, t, \alpha)$. This resolution function accounts for the position resolution of the PSD, the $\lambda_n$ 
resolution of the ToF, 
as well as the divergence distribution of the beam. This product of terms is multiplied by a scaling factor $N$ and a uniform background term $B$ is added such that the final fit function takes the form 
 $f(x_i, t_i, \vec{\theta}) = N \mathcal{R}(x_i - x', \lambda_n(t_i) - \lambda_N', \alpha-\alpha')*n(\lambda_n') j(x', \lambda_n', \alpha')' + B$.
The vector $\vec{\theta}$ encodes all of the fit parameters of the function. Those include the mirror radius, mirror angular size, mirror optical potential, beam incidence angle, chopper-mirror distance, mirror-detector distance, timing offset, beam position offset, normalization, background, spin-independent shift, and the spin-dependent shift.

To account for a spin-dependent $\lambda_n$ 
shift, the two spin data sets are fit simultaneously with the fit function as above but with a spin-dependent $\lambda_n$ 
shift as defined in Eq. \ref{eq: Wavelength Shift} but with $   \delta g_{\mathrm{Effective}}^\pm = g_0 \pm \delta g_{\mathrm{WGS}}/2$,
 where $g_0$ is an overall shift to account for hypothetical spin-independent shifts and $\delta g_{\mathrm{WGS}}$ is the spin dependent shift of interest. Overall shifts could arise from spin-independent systematic uncertainties, like imprecise knowledge of the radius of curvature or angular size of the mirror.

The fitting is done in a Bayesian framework, that is, the negative log of the posterior distribution
\begin{equation}
    p(\vec{\theta}^+,\vec{\theta}^-|\mathcal{D}^+,\mathcal{D}^-) = \mathcal{L}(\mathcal{D}^+,\mathcal{D}^-|\vec{\theta}^+,\vec{\theta}^-) \pi(\vec{\theta})
    \label{eq: posterior}
\end{equation}
is minimized. In Eq. (\ref{eq: posterior}), $\vec{\theta}^\pm$ are the fit parameters and $\mathcal{D}^\pm$ data sets for spin-up and spin-down, respectively. The likelihood is denoted as $\mathcal{L}$ and the prior distributions as $\pi(\vec{\theta})$. The likelihood is defined as 
\begin{equation}
    \begin{split}
       \mathcal{L}(\mathcal{D}^+,\mathcal{D}^-|\vec{\theta}^+,\vec{\theta}^-) &= \mathcal{L}(\mathcal{D}^+|\vec{\theta}^+)\mathcal{L}(\mathcal{D}^-|\vec{\theta}^-) \\
       & = \prod_i \frac{f(x_i, t_i, \vec{\theta}^+)^{n^+_i} \exp{(-f(x_i, t_i, \vec{\theta}^+))}}{n^+_i!} \\ & \times \frac{f(x_i, t_i, \vec{\theta}^-)^{n^-_i} \exp{(-f(x_i, t_i, \vec{\theta}^-))}}{n^-_i!}
    \end{split}
\end{equation}
and the priors are all taken to be Gaussian distributions, that is, $\pi(\vec{\theta})$ should be interpreted as    $\pi(\vec{\theta}) = \prod_j \frac{1}{\sigma_j \sqrt{2\pi}} \exp{\left(- \frac{(\theta_j-\mu_j)^2}{2\sigma_j^2} \right)}$,
where $\mu_j$ and $\sigma_j$ are our prior estimate of $\theta_j$ and our uncertainty of that estimate. These values are determined by the precision of the D17 experimental set-up, as well as auxiliary systematic measurements made specifically of the WGS 
set-up. The details of these studies are the subject of a dedicated article to be published. 

The minimization of $-\log{p(\vec{\theta}^+,\vec{\theta}^-|\mathcal{D}^+,\mathcal{D}^-)}$ is done via Newton's method. The minimum $\vec{\hat{\theta}}$ is taken as our fit result and the inverse of the Hessian at this minimum is taken as our covariance. Here we approximate the posterior near its maximum using Laplace's approximation.
 
\begin{figure}[h!]
  \centering
  \includegraphics[width=1.1\linewidth]{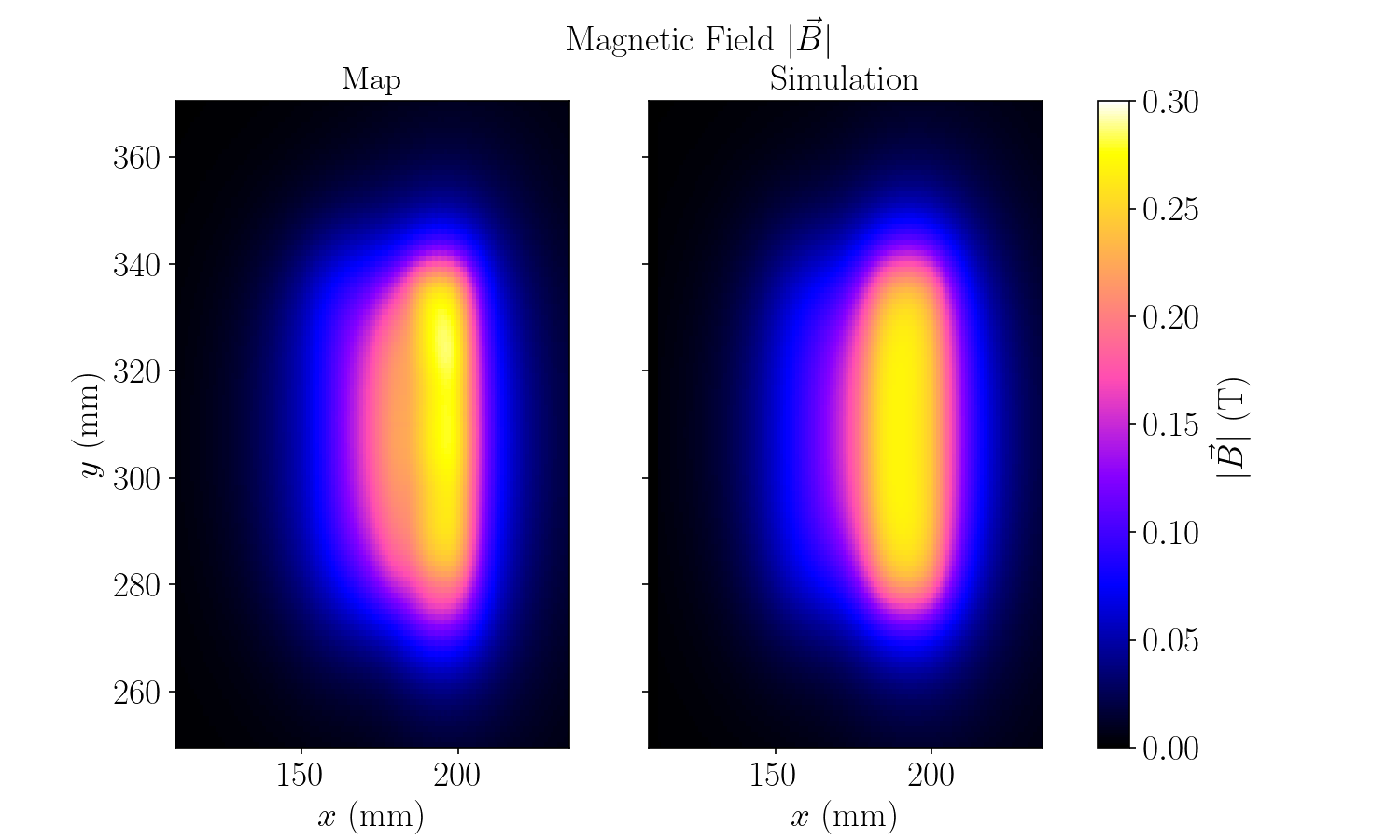}
  \caption{Results of magnetic field map for the magnet in 2025 (left) vs simulations done with Radia (right). The maximum deviation between simulation and the map is $\sim10\%$}
  \label{B_fields}
\end{figure}

\emph{Uncertainty of the magnetic field gradient}. 
The magnetic field and the gradient generated by this magnetic field was designed using the Radia software package \cite{ochubar2015radia} in Mathematica. Comparisons of this simulation with a field map can be found in Fig. \ref{B_fields}.
The magnet upstream of the mirror was added   
to amplify the magnetic shift. 

The magnet inducing the shift comprised three sets of NdFeB magnets that were stacked in the orientation shown in Fig. \ref{fig:setup}. They were aligned in an Aluminum channel and fixed to the holder of the curved mirror so that the bottom of the magnet was \SI{8}{\milli\meter} above the flat edge of the mirror. The magnet was centered on the \SI{40}{\milli\meter} beam. Since each magnet has a depth of \SI{20}{\milli\meter}, the 3 sets of magnet together has a depth of \SI{60}{\milli\meter}, enabling us to treat the gradient as uniform along the beam width. 

The magnetic field was mapped with a robotic arm with a positioning precision of 100 $\mathrm{\mu}$m; an example of the mapping is shown in Fig. \ref{B_fields}. The measurement was repeated for several distances of the probe from the surface of the channel, to ensure that the B-field was well characterized in three dimensions. We conclude from these measurements that the magnetic field strength and gradient on the mirror surface is as assumed in our simulations, and the magnets were therefore well constructed.

The theoretical uncertainty of the shift was determined by propagating the precision of our knowledge of the magnet construction through the Radia simulation and seeing its effects on $\delta g_{\mathrm{WGS}}$. We changed parameters like the strength and direction of the magnetization vector of each NdFeB magnet, as well as their positions with respect to each other and to the mirror. The width of the beam is considered by averaging the value of $\delta g_{\mathrm{WGS}}$ over various entrance positions of $n$ into the gallery. The results of this analysis are in Table \ref{Table: Systematic Effects}. 
\begin{table}[h]
\centering
\begin{tabular}{lc}
\textbf{Parameter} &
\textbf{$\sigma_{\delta g_{\mathrm{WGS}}}$ (\SI{}{\meter\per\second\squared})} \\ \hline
\hline 
NdFeB Remanence  & $0.13\times 10^3$ \\
Magnet Vertical Position & $0.16\times 10^3$\\ 
\end{tabular}
\caption{We only list the most important contributions to the uncertainty of the theoretical prediction. The uncertainty of the remanence field ($\sim 10 \%$) of each individual NdFeB magnet in our magnet is propagated through the simulation of $\delta g$ and added in quadrature, resulting in the first row of the table. The uncertainty of the height of the magnet with respect to the flat edge of the mirror (taken to be $0.5$ mm) results in the second row.}
\label{Table: Systematic Effects}
\end{table}

\emph{Lessons Learnt.}
In a previous version of this experiment, 
we observed that the $n$ count rate seemed to follow a trend correlated with the temperature in the $n$ 
guide hall \cite{Nesvizhevsky2025MagneticShift}. We did not see a similar change in the background count rate, indicating that only the collimated part of the beam was affected. Therefore 
we constrained the analysis to shorter time
spans to evaluate the effect on the shift value drift. We observed that when analyzing consequent 10x15 minutes batches of data, we could see also a drift in the calculated shift value. 

To improve the $n$ count stability, we opened the $s_3$ slit 4-times wider in the latest measurement. 
A thermal misalignment would therefore have a less significant effect, as the beam would have a larger margin before it would be cut out from the ideal sample-hit position. To monitor if the sample holder changes position with a change in temperature, we mounted a laser pointer on the sample stage, which hit a photodetector fixed on the concrete wall of the experimental area several meters away. We did not observe significant changes of the laser beam position, from which we conclude that the sample was very stably positioned. 

There was an increase in the background count rate with the closing of the neighboring instrument's shutter. We monitored our count rate along with the closing and opening times of our neighbors shutter, and we were able to use 
the majority of the data with the proper background treatment. In the overall count rate during the beamtime, two events were registered, where count rates were significantly below what we would expect. As no change in laser position was sensed at these times, we assume that these are just unphysical outliers, and neglect them in the analysis.

\begin{figure}
   \centering
       \includegraphics[width=0.95\linewidth]{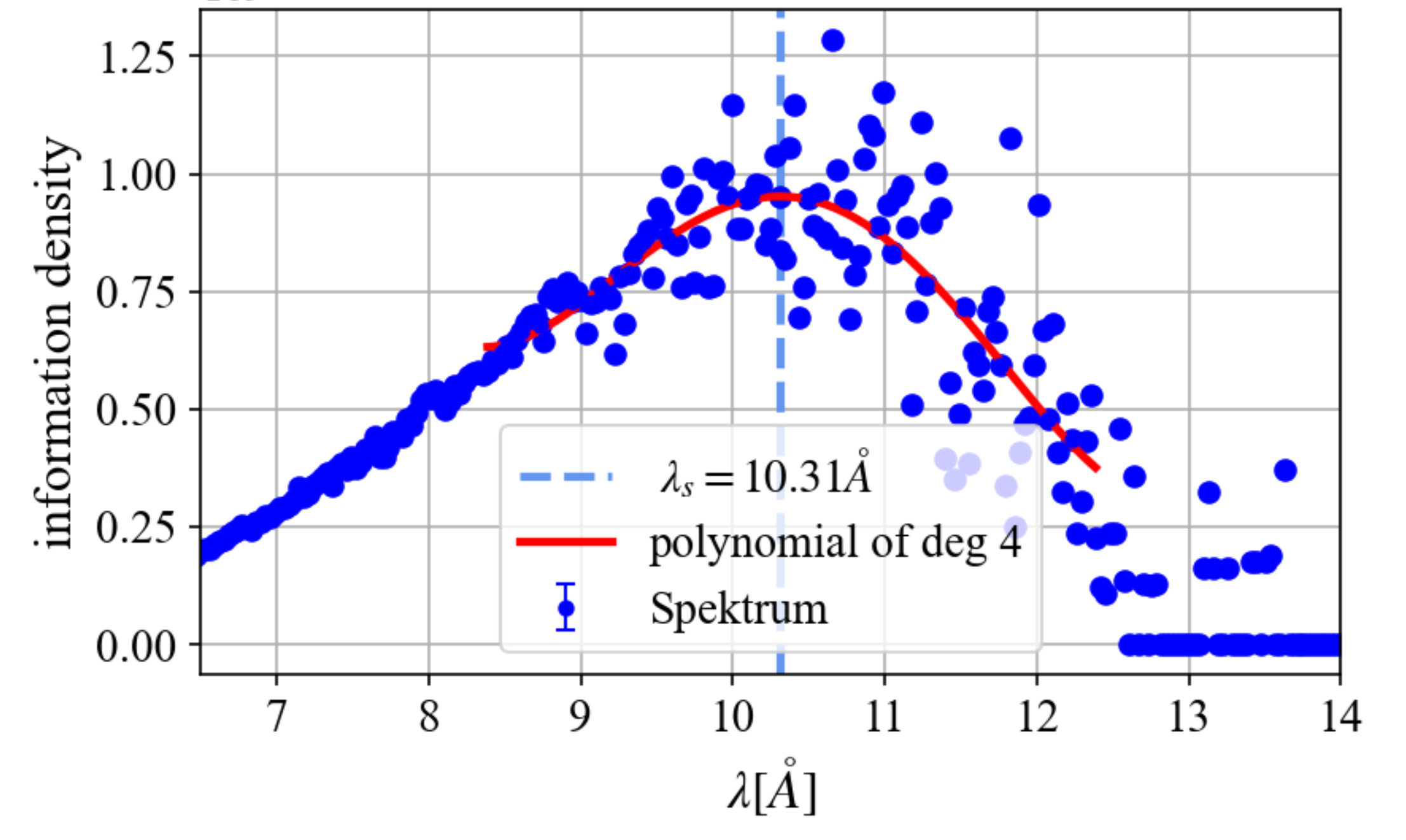}
    \caption{The Fisher information density, as well as its 
    4$^{th}$ degree polynomial fit. 
    The sensitivity is maximal at 
    $\lambda_n \sim 10.3$~\AA. 
    }
    \label{fig:sensitivity}
\end{figure}

\emph{Estimation of the maximal sensitivity $\lambda_n$}. The $n$ count $P(\lambda_n)$ is evaluated from the measured WGS signal at $40^o$. 
Shifts are scaling as $\propto \delta_g \cdot \lambda_n^3$ so we define our expected sensitivity 
as:$ h(\lambda_n) \propto \lambda_n^3 P(\lambda_n)$. 
For the Poissonian variance $\sigma^2(\lambda_n) \approx P(\lambda_n)$, the Fisher information is $I(\lambda_n
) \approx \int \frac{\left( \partial(\delta_g \lambda_n^3)/\partial \delta_g\right)^2}{\sigma^2(\lambda_n)}\mathrm{d\lambda_n} = \int \lambda_n^6 P(\lambda_n) 
\mathrm{d\lambda_n}$.
Fig. ~\ref{fig:sensitivity} shows the Fisher information density curve and the fit to extract the $\lambda_n$ value of the maximum sensitivity. 

\end{document}